\documentclass[fleqn,11pt]{ices05}
\usepackage{graphicx}

\begin{document}
{\small

\title{Computation of Electrostatic Field near Three-Dimensional Corners and
Edges}

\author{N.Majumdar$^1$, S.Mukhopadhyay\footnote{Saha Institute of Nuclear
		Physics, 1/AF, Sector 1, Bidhannagar, Kolkata 700064, West Bengal,
		India, e-mail: nayana.majumdar@saha.ac.in}
	}
}

\abstract{Theoretically, the electric field becomes infinite at corners of two
and three dimensions and edges of three dimensions. Conventional finite-element
and boundary element methods do not yield satisfactory results at close
proximity to these singular locations. In this
paper, we describe the application of a fast and accurate BEM solver (which uses
exact analytic expressions to compute the effect of source distributions on flat
surfaces) to compute the electric field near three-dimensional corners and
edges.
Results have been obtained for distances as close as 1$\mu m$ near the corner /
edge and good agreement has been observed between the present results and
existing analytical solutions.}

\section{Introduction}
Accurate computation of electric field near corners and edges is important in
many applications related to science and engineering. While the generic problem
is important even in branches like fluid and solid mechanics, the specific
electrostatic problem is very important in the modeling and simulation of
micro electro-mechanical systems (MEMS), electric probes and antennas, to name a
few. While it is true that the ideal corner / edge does not exist in nature, the
singularity being smoothed by rounding corners, sharp increase in charge density
does occur near these geometries. Since the electric field is proportional to
the square of the charge density, it is very important to estimate the charge
density and the resulting variation of electric field in their vicinity.

There have been many attempts at solving the above problem using the
finite-element (FEM) and the boundary element method (BEM). However, because of
the nature of the
problem, significant modifications to the basic method needed to be carried out
to handle the boundary singularities. On the simpler side, special mesh
refinement schemes have been used while on the more sophisticated side, the
form of local asymptotic expansions (which may often be found) have been used.
Very effective FEM solvers have been developed \cite{Babuska}
which calculate singular coefficients by post-processing the
numerical solution. These solvers improve the solution by refining the mesh and
changing the degree of piecewise polynomials. Similarly, several accurate
BEM solvers have been developed \cite{Igarashi}, \cite{Elliotis}.
For example, in the singular function boundary integral method,
the singular coefficients are calculated directly and the solution is
approximated by the leading terms of the local asymptotic solution.

In this paper, we present a solution to the corner / edge problem by using
a recently developed three-dimensional BEM solver \cite{Mukhopadhyay}. This
solver uses exact
analytic expressions for computing the influence of charge evenly distributed
on discretized flat elements on the boundary. Through the use of these
closed-form expressions, the solver avoids one of the most important
approximations
of the BEM, namely, the assumption that the charge on a boundary element is
concentrated at one point (normally the centroid) of the element. As a result,
the computed capacitance matrix is very accurate and the solver is able to
handle difficult
geometries and charge distributions with relative ease. The solver has been
used to solve the classic problem of two planes intersecting at various angles.
Exact solution to this problem exists \cite{Jackson} and our results have been
compared with the exact results. The comparison is very good even as close as
1 $\mu m$ to the corner / edge. Especially important is the fact that the
solver produces quite accurate results even for the case of an edge. It has also
been possible to reproduce the dependence of the strength of the electric field
as a function of the distance from the geometric singularity. All the
calculations have been carried out in three dimensions and, here, we also
present the variation of the electric field along the corner or edge. It is
observed that the electric field retains its mid-plane value for much of the
distance along its length, but increases significantly within 20\% of the axial
ends. It may be noted here
that for these calculations, only algebraic mesh refinement near the edge was
applied and no other sophisticated techniques such as those mentioned above
were applied. This fact made the development of the solver and its application
free of mathematical and computational complications. Since corners and
edges play an important role in many electro-mechanical structures, the solver
can be used to study the electrostatic properties of such geometries.

\section{Theory}
We have considered the geometry as presented in \cite{Jackson} in which two
conducting planes intersect each other at an angle $\beta$. The planes are
assumed to be held at a given potential. A circular cylinder is also included that just
encloses the two intersecting plane, has its center at the intersection point
and is held at zero potential. The general solution in the polar coordinate
system ($\rho$, $\phi$) for the potential ($\Phi$) close to
the origin in this problem has been shown to be
\begin{equation}
\label{eq:Potential}
\Phi(\rho, \phi) = V\,
+\, {\sum_{m=1}^{\infty}}\, a_m\, \rho^{m \pi / \beta}\,sin(m \pi \phi/\beta)
\end{equation}
where the coefficients $a_m$ depend on the potential remote from the corner at
$\rho = 0$ and $V$ represents the boundary condition for $\Phi$ for all $\rho
\geq 0$ when $\phi = 0$ and $\phi = \beta$. In the present case where a circular
cylinder just encloses the two plates, the problem of finding out $a_m$
reduces to a basic fourier series problem with a well known solution
\begin{equation}
\label{eq:Coeff}
a_m\, =\, \frac{4}{m \pi}\,\,\, for\, odd\, m
\end{equation}
It may be noted here that the series in (\ref{eq:Potential}) involves
positive powers of $\rho^{\pi/\beta}$, and, thus, close to the origin (i.e., for
small $\rho$), only the first term in the series will be important. The electric
field components ($E_{\rho}, E_{\phi}$) are
\begin{equation}
\label{eq:Efield_rho}
E_{\rho}(\rho,\phi) =
-\frac{\pi}{\beta}
{\sum_{m=1}^{\infty}}\, a_m\, \rho^{(m \pi/\beta) - 1} sin(m \pi \phi / \beta)
\end{equation}
\begin{equation}
\label{eq:Efield_phi}
E_{\phi}(\rho,\phi) =
-\frac{\pi}{\beta}
{\sum_{m=1}^{\infty}}\, a_m\, \rho^{(m \pi/\beta) - 1} cos(m \pi \phi / \beta)
\end{equation}
Thus, the field components near $\rho = 0$ vary with
distance as $\rho^{(\pi/\beta) - 1}$ and this fact is expected to be reflected
in a correct numerical solution as well.

\section{Numerical Solution}
While the above theoretical solution is a two-dimensional one, we have used the
BEM solver to compute a three-dimensional version of the above problem.
In order to reproduce the two-dimensional behavior at the mid-plane, we have
made the axial length of the system sufficiently long, viz., 10 times the
radius of the cylinder. The radius of the cylinder has been fixed at one meter,
while the length of the intersecting flat plates has been made a micron shorter
than the radius. The length of the plate has been kept smaller than the radius
of the cylinder to avoid the absurd situation of having two values of the
voltage at the same spatial point. We believe that it has been possible to
maintain
such a small gap as 1 $\mu m$ between the circular cylinder and the flat plates
without facing numerical problems because the BEM solver we have used computes
the capacity matrix using analytic expressions which calculate accurate values
of potential and electric field at extremely close distances from the singular
surfaces \cite{Mukhopadhyay}.

The cylinder has been discretized uniformly in the angular and axial directions.
The flat plates have also been uniformly discretized in the axial direction.
In the radial direction, however, the flat plate elements have been made
successively smaller towards the edges using a simple algebraic equation in
order to take care of the fact that the surface charge density increases
considerably near the edges. The
electric field has been computed for various values of $\beta$ ($1.25 \pi,\,
1.5 \pi,\, 1.75 \pi\, and\, 2 \pi$),
referred to as cases 1, 2, 3 and 4 respectively in the following section.

\section{Results}
\begin{figure}
\vspace{-0.5in}
\begin{center}
\includegraphics[height=2in,width=4in]{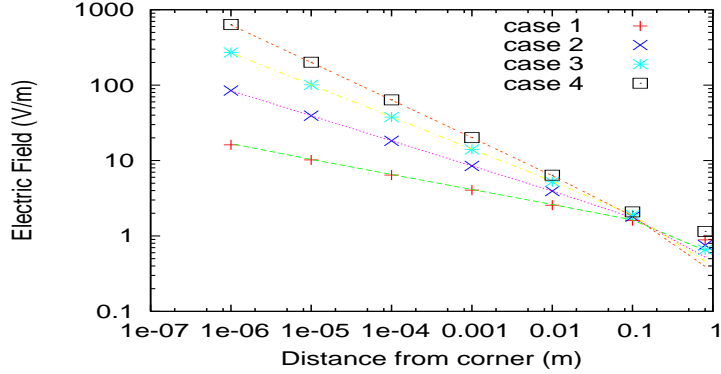}
\caption{\label{fig:ef1} Comparison of electric field values}
\end{center}
\end{figure}

In Figure\ref{fig:ef1}, we have presented a comparison of the variation of the
electric field as a function of the distance from the edge as found from the
analytical solution (\ref{eq:Efield_rho}) and the BEM solver. Computations have
been
carried out upto a distance of 1 $\mu m$ from the edge and to properly
represent the sharp increase in field, logarithmic scales have been
used. The computed electric field is found to be in remarkable agreement for
all values of $\beta$. There is a small disagreement between the two results
only at the point closest
to the corner / edge ($\approx 1.3\%$). At present, we are in the process of
improving the BEM solver so that this error can be minimized.

\begin{figure}
\vspace{-0.5in}
\begin{center}
\includegraphics[height=2in,width=4in]{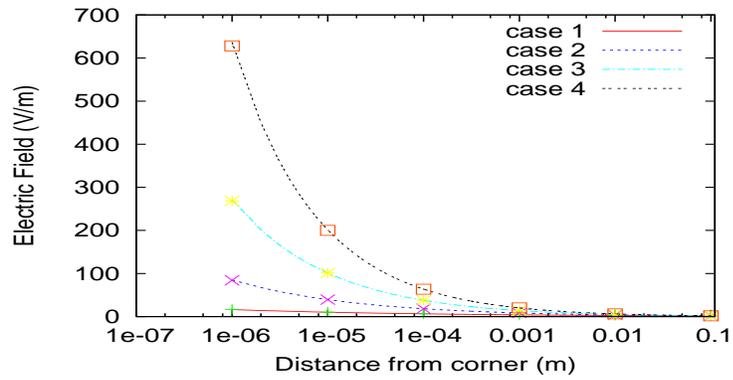}
\caption{\label{fig:ef2} Dependence of electric field as a function of distance}
\end{center}
\end{figure}

Although, the distance dependence of the electric field is expected to
match with the theoretical prediction because the computed and the theoretical
estimates are found to match closely, in Figure \ref{fig:ef2}, we have plotted
curves corresponding to $\rho^{\pi / \beta\, -\, 1}$ and compared the computed
values as points against the curves. Thus, we can confirm that the computed
electric field obeys the $\rho^{(\pi/\beta)\, -\, 1}$ relation quite accurately
near the corner / edge. Here, to emphasize and visualize the sharpness of the
increase of electric field near the singular geometry, only the distance has
been drawn using the logscale.

\begin{figure}
\vspace{-0.5in}
\begin{center}
\includegraphics[height=2in,width=4in]{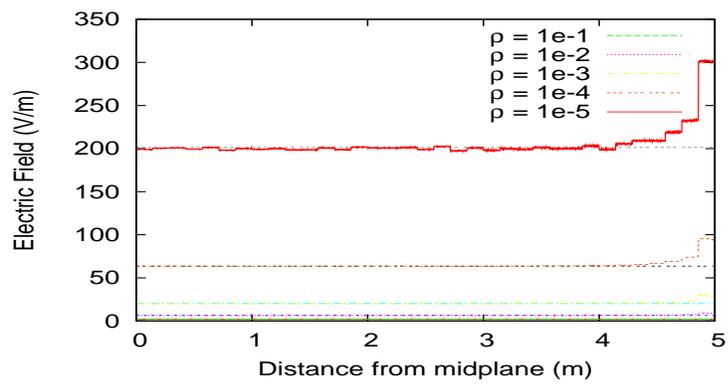}
\caption{\label{fig:ef3d} Variation of electric field along the axis}
\end{center}
\end{figure}

In Figure\ref{fig:ef3d}, we have presented the variation of the electric field
at various distances from the corner / edge along the length of the geometry.
While the field is found to increase considerably towards the end, for most
of the length ($\approx 80\%)$, the two-dimensional value seems to be a good
approximation for the real value. However, the deviation (as large as
$\approx 50\%$) from the analytic solution towards the
front and back ends of the geometry should be taken into consideration while
designing real-life applications. At close distances near the singular location,
oscillation in the electric field value is observed. Initial study indicates
that the source of this oscillation is possibly the large size of the boundary
elements in the axial direction.

In order to facilitate numerical comparison, we have presented the values of the
electric field in the mid-plane as obtained from the exact solution and the BEM
solver in Table \ref{table:table1} for the three dimensional edge. We have
chosen this case since it is known to be the most difficult among all those
considered in this work. It is clear from the table that the present solver
does indeed yield very accurate results.
\begin{table}
\vspace{-0.5in}
\centering
\begin{tabular}{| l | c | c | c |}
\hline
Distance & Exact & Computed & Error (\%) \\
\hline
0.1 & 1.83015317 & 1.829997 & 0.0085332 \\
\hline
0.01 & 6.303166063 & 6.302387 & 0.012359868 \\
\hline
0.001 & 20.11157327 & 20.103534 & 0.039973352 \\
\hline
0.0001 & 63.65561168 & 63.548305 & 0.168573794 \\
\hline
0.00001 & 201.31488353 & 199.998096 & 0.654093481 \\
\hline
0.000001 & 636.6191357 & 628.098310 & 1.33844951 \\
\hline
\end{tabular}
\caption{\label{table:table1}Comparison of electric field values}
\end{table}

\section{Conclusions}
A fast and accurate BEM solver has been used to solve the complex problem of
finding the electrostatic force field for three dimensional corners and
edges. Accurate solutions have been obtained upto distances very close
singular geometry. The two dimensional analytic solution has been found to be
valid for a large part of the geometry, but near the axial ends, the difference
has been found to be significant. The results and the BEM solver is expected
to be very useful in solving critical problems associated with the design of
MEMS, integrated circuits, electric probes etc. Since these problems are
generic in nature, the solver should be important for analysis of problems
related to other fields of science and technology.

\reference{
\item \label{Babuska}
I Babuska, B Guo (1986):
"The h-p version of the finite element method Part I: the basic
approximation results", \textit{Computational Mechanics}, 33, 21-41.

\item \label{Igarashi}
H Igarashi, T Honma (1996):
"A boundary element method for potential fields with corner singularities",
\textit{Appl. Math. Modelling}, 20, 847-852.

\item \label{Elliotis}
M Elliotis, G C Georgiou, C Xenophontos (2002):
"The solution of a Laplacian problem over an L-shaped domain with a singular
boundary integral method",
\textit{Commun. Numer. Meth. Eng.}, 18, 213-222.

\item \label{Mukhopadhyay}
S Mukhopadhyay, N Majumdar (2005):
"Development of a BEM solver using exact expressions for computing the
influence of singularity distributions",
submitted to the \textit{International Conference on Computational \&
Experimental Engineering and Sciences} to be held at Indian Institute of
Technology, Chennai from 1 to 6 December, 2005.

\item \label{Jackson}
J D Jackson (1988):
"Classical Electrodynamics", $2^{nd}$ edition,
\textit{Wiley Eastern Limited}, (1988), 75-78.
}
\end{document}